\documentclass[epjST]{svjour}

\usepackage{subfigure}
\usepackage{graphicx}
\usepackage [autostyle, english = american]{csquotes}
\MakeOuterQuote{"}
\usepackage[final]{changes}
\begin{document}

\title{An overview of memristive cryptography}
%\subtitle{A Review of the State of the Art}
\author{Alex Pappachen James\inst{1}\fnmsep\thanks{\email{apj@ieee.org}} }
\institute{Nazarbayev University}
\abstract{ 
Smaller, smarter and faster edge devices in the Internet of things era demands secure data analysis and transmission under resource constraints of hardware architecture. Lightweight cryptography on edge hardware is an emerging topic that is essential to ensure data security in near-sensor computing systems such as mobiles, drones, smart cameras and wearables.
In this article, the current state of memristive cryptography is placed in context of lightweight hardware cryptography. The paper provides a brief overview of the traditional hardware lightweight cryptography and cryptanalysis approaches. The contrast for memristive cryptography with respect to traditional approaches is evident through this article, and need to develop a more concrete approach to developing memristive cryptanalysis to test memristive cryptographic approaches is highlighted.
} %end of abstract
\maketitle
\section{Introduction}
\label{intro}

The penetration of internet into every aspect of our lives also brings with it several challenges related to data security and privacy \cite{zhang2014iot,dorri2017blockchain}. Manipulation and misuse of data can have a substantial influence in the way we perceive and view our world \cite{cache2007hacking,warren2006social}. The cryptography \cite{katz1996handbook} studies deals with encryption and decryption, while cryptanalysis \cite{stamp2007applied} deals with the techniques to break encrypted systems. The existing cryptography methods \cite{balasubramanian2018recent} are largely based on mathematical theories designed for computational hardness, with an aim to make it difficult for an adversary to break into such systems.

The \replaced{vulnerability}{venerability} of the encryption techniques are often exposed through various side channel attacks \cite{zhou2005side,brier2002weierstrass,joy2011side} and through high performance computing tools. It is expected that with future technologies such as quantum computing \cite{sergienko2018quantum} can introduce massive parallelism that can make most of the encryption techniques look very weak. Given the challenges are significant in the years ahead, it is only important to address this topic in a hardware perspective in view of the challenges ahead with post-quantum cryptography era \cite{buchmann2018postquantum}. The exclusive need to ensure secure data processing in edge devices with internet requires high speed and low power offered by hardware circuits that is not beatable by the existing software only counterparts \cite{damaj2018analysis}. 

The hardware based cryptography \cite{rajagopalan2012survey,de2007high} has been in use for several decades, as it offers a faster and efficient way to generate keys and random numbers. In addition, embedded in reconfigurable chips, or that in ASIC, it is practically difficult to decode the logic or implement various side channel attacks \cite{el2015survey}. The dynamic nature of such keys makes it extremely hard to break. With advancement \replaced{of}{} wearable and internet of things devices, it becomes even more important to provide on-chip solutions, that are area and power efficient \cite{pantelopoulos2010survey}. The ability to have low power solutions are important as many of these wearable works on limited battery capacity, and often require secure data transmission \cite{ometov2016feasibility}. The implementation of the existing algorithmic only solutions are not efficient in such situations, and nano-electronic solutions become more viable \replaced{}{solution}.

In the last decade, there has been a substantial push towards more than Moore's era technologies \cite{huff2008into,williams2017s1}, with focus on emerging devices for non-traditional computing architectures and systems. This is required to overcome the limitations imposed by device scaling \cite{kahng2010scaling} and the rapid need to have higher computational capabilities for edge devices \cite{krestinskaya2019neuro}.  In this review, we present the overall developments in the hardware based cryptography with specific focus on the use of memristor devices and networks. The importance of this topic lay in the intersection of memristor as \replaced {an}{a} effective device used for chaotic system, having ability to switch states, and having interesting properties that resemble the generalisation functions of a neuron and its networks.

\added{The paper is organised into five sections: section 2 provides an introduction to memristors and memristor networks, section 3 provides background on lightweight cryptography, section 4 builds on the previous section to introduce memristor cryptography and section 5 concludes the paper. }

\section{Memristor networks}

The memristor \replaced{(Fig. 1(a,b))}{} remained as an \replaced{elusive}{illusive} circuit element for several decades, until the claims of this missing circuit element being found was proposed in 2008 \cite{strukov2008missing}. After this, there have been a surge of memory devices that is deemed fit into the broad category of memristors. They find applications in as non-volatile memory, modelling neural networks, chaotic circuits, signal processing, and cryptography. In several of these applications, the most popular memristor circuit configurations is that of a memristor crossbar configuration  \replaced{(Fig. 1(c))}{}, which can be used for memory array, and for dot-product computations. 

\begin{figure}[ht]
\centering
\subfigure[]{
    \includegraphics[width=80mm]{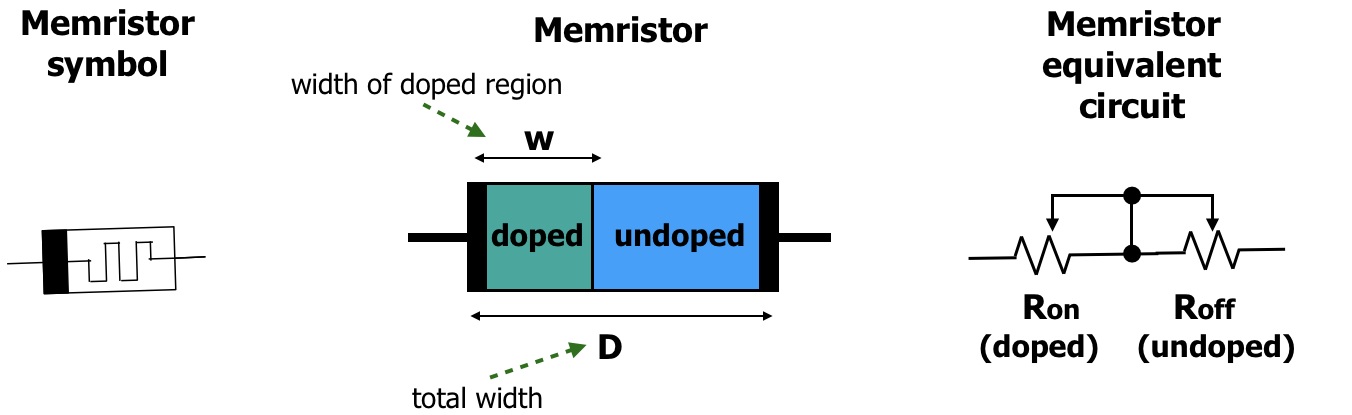}
    \label{fig:subfig1}}
\subfigure[]{
     \includegraphics[width=50mm]{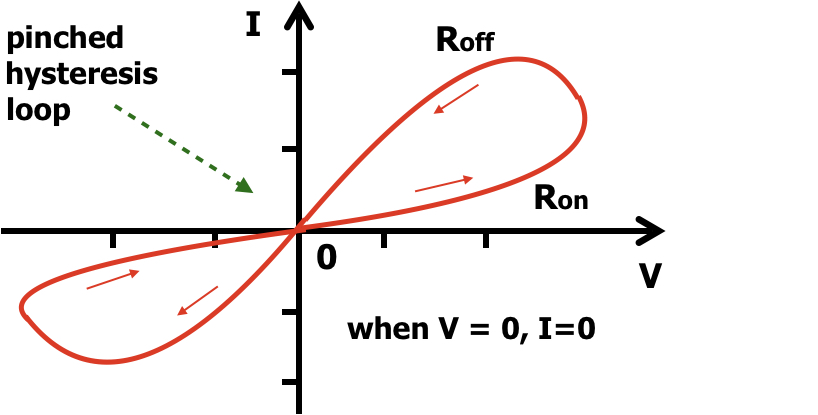}
    \label{fig:subfig2}}
\subfigure[]{
     \includegraphics[width=50mm]{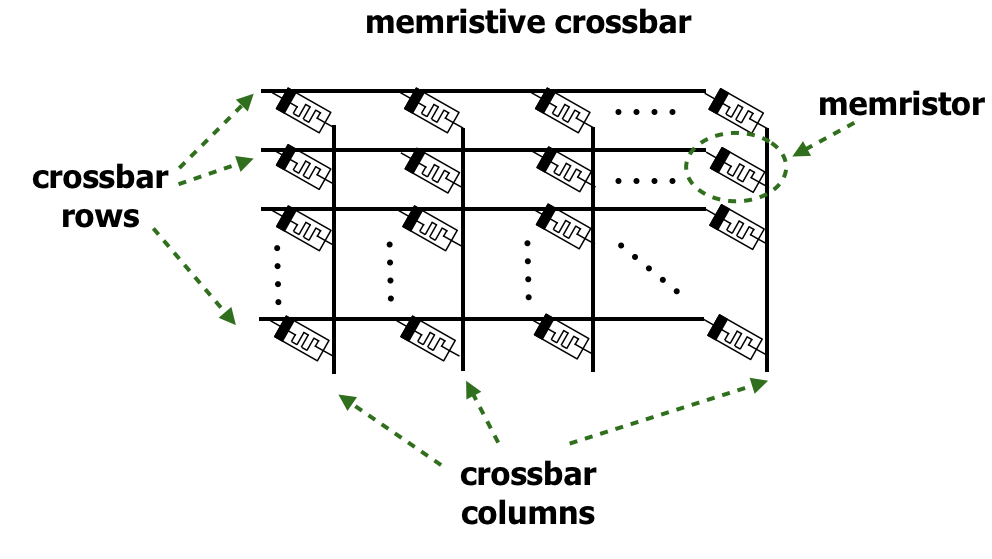}
    \label{fig:subfig3}}
\caption{ \subref{fig:subfig1} shows an example memristor symbol and its equivalent representation, \subref{fig:subfig2} shows the IV characteristics and \subref{fig:subfig3} shows the memristor array in crossbar.}
\label{fig:subfigureExample}
\end{figure}

\subsection{Memristor in a nutshell}
The memristor is considered as a fourth fundamental circuit element \cite{strukov2008missing,chua1971memristor}. There has been arguments in the recent past for and against this assertion \cite{vongehr2015missing,abraham2018case}. Nonetheless, there are several useful behavioural properties that makes memristor practically a very useful circuit element \cite{joglekar2009elusive,ho2011dynamical,corinto2018memristor}. In a recent paper, five enigmas of non-volatile  memristor device theory \cite{chua2018five} were proposed and proved:

\begin{enumerate}
\item Enigma 1: \textit{All non-volatile memristors have continuum memories.}
\item Enigma 2: \textit{Conductance of all non-volatile memristors can be tuned by applying single voltage pulses.}
\item Enigma 3: \textit{Faster switching can always be achieved by increasing the pulse amplitude.}
\item Enigma 4: \textit{Periodic unipolar input gives non-periodic finger-like multi-prong-pinched hysteresis loops.}
\item Enigma 5: \textit{DC VI curves of non-volatile memristors are fakes.}
\end{enumerate}

These enigmas provide the summary of what we know today about idealistic memristor device. In fact, modelling of any realistic memristor with high accuracy is very challenging task. The underlying reason for this is largely due to the material characteristics of the devices that vary significantly from one device to another. The device level variability issues for large majority of memristor devices is still not resolved to completely validate with mathematical models with accuracy's similar to standard CMOS technology. Any simulation of memristors without variability analysis is incomplete and does not reflect a realistic implementation. \replaced{In contrast, the variability of memristor specific to a manufacturing process is often useful for cryptographic application such as to generate random keys and physically unclonable functions.}{}  

\subsection{Crossbar arrays}

The crossbar latch  \cite{kuekes2005crossbar} is one of the memristor array configuration that was shown to be useful for implementing various digital logic operations. The memristor crossbar array architecture can also be used for writing and reading conductance values of the memristor making it useful as a memory array. The crossbar architecture can also be used for building analog neural computing units.

In a crossbar arrangement of memristors \cite{mouttet2008proposal,mouttet2007programmable,vontobel2009writing}, the inputs are applied to the rows as voltage signals and outputs are read as current signals. The current output is a weighted sum of input voltage, where the weights corresponds to the memristor conductance\cite{kim2011functional}. Mathematically, this is equivalent to a dot product operator which is required for weighted summation of inputs in each neural network layer \cite{zhang2018neuromorphic,krestinskaya2018analog}. The two-terminal memristor devices are area efficient, and can help accelerate neural network computations at high speed and low power. The memristor crossbar can also be used as a regular memory array, with each memristor in the network is capable to be programmed to several discrete resistive states \cite{adam20173,lu2011two}.

The variability of the memristor states from device to device under the same conditions and constraints, often is considered as a challenge for having stable analog memory \cite{irmanova2018neuron,stathopoulos2017multibit}. This makes the use of memristor as an analog memory in large crossbar array not practical, however, as a discrete state device and as a binary state device, they can be used effectively, in small arrays. The crossbar also suffers from sneak path problems,  parasitic resistors, and wire resistors, that can further limit the large scaling of crossbar that can be implemented today \cite{li2018analogue,krestinskaya2019memristive}.

\section{Lightweight cryptography}

\subsection{Cryptograpic methods}
Lightweight cryptography \cite{eisenbarth2007survey} works between the trade-offs of security, cost, and performance, and is focused at devices and systems on edge. The increase in internet connected devices, requires to build smarter systems that are secure using low-cost hardware solutions. The symmetric and
asymmetric ciphers are essentially a major topic of study in hardware cryptography, each having a different set of applications. Hardware for asymmetric ciphers are more complex than symmetric ones, and consume more area on chip and power. For example, in terms of computational complexity, symmetric cipher such as the Advanced Encryption Standard (AES) \cite{daemen2013design} algorithm is about 1000 much faster than an optimised elliptic-curve cryptography \cite{hankerson2011elliptic} that is an asymmetric algorithm. 

There exists several hardware implementations  of  ciphers such as  Hight\cite{hong2006hight}, Clefia\cite{shirai2007128}, DESXL \cite{panasenko2011lightweight}, DESL, SEA\cite{moosavi2015sea}, Hummingbird\cite{engels2010hummingbird}, PRESENT\cite{poschmann2009lightweight}, PRINTcipher\cite{knudsen2010printcipher}, mCrypton\cite{lim2005mcrypton}, KLEIN\cite{gong2011klein}, TWINE\cite{tomoyasu2012twine}, SIMON \cite{beaulieu2015simon}, SPECK\cite{beaulieu2015simon}, PRINCE\cite{borghoff2012prince}, PRIDE\cite{albrecht2014block}, LBLOCK\cite{wu2011lblock}, MIBS\cite{izadi2009mibs}, Puffin \cite{cheng2008puffin}, ESF\cite{tripathy2013esf}, Piccolo\cite{shibutani2011piccolo}, Khudra\cite{kolay2014khudra} etc in use today, making this an emerging topic of study for edge devices. In addition, there are also several requirements for AEAD\cite{struik2013aead} and hash functions\cite{balasch2012compact}, for lightweight cryptography, such as, they should be useful for short messages, optimised for resource constraint hardware, efficient key preprocessing, apply to different platforms, and have low power/energy/latency. The resource constraints also means that there are higher 'risk' in design, lower security margins, and few number of components that can be targets of attacks. 

\subsection{Cryptanalysis methods}
The analysis of the cryptography algorithms in general is known as cryptanalysis\cite{schneier2000self}, and is an essential aspect of testing the reliability of the cryptography system for practical use. The major classes of attacks\cite{sun2015links} can be classified as based on impossible differentials, guess and determine, and that are dedicated for a given method. In classical differential attack\cite{karaklajic2013hardware}, the difference two outputs relative to the difference in pain text is tracked. In a truncated differential attacks\cite{knudsen1994truncated}, changes to only part of the differences are predicted. The impossible differential attacks\cite{kim2003impossible} on the other hand uses a differential with probability 0. The  Miss-in-the-middle \cite{biham1999miss} improves over this, and by extending such approaches to forward and backward can give information of key bits\cite{wang2014cryptanalysis}. 

There have been numerous improvements to impossible differential attacks such as multiple impossible differentials, choosing the correctly the changes, state-test techniques, and improving the estimate of pairs \cite{boura2014scrutinizing}. The example applications of impossible differential attacks includes a  best attack on CLEFIA with 13 rounds\cite{mala2011impossible}, improved best attacks for Camellia\cite{wu2007impossible}, AES attacks comparable with best mitm ones in 7 rounds \cite{mala2010improved} and LBlock with reduced rounds \cite{karakocc2012impossible}.

The Meet-in-the-middle\cite{nishimura1990probability} attack is relatively an old approach, that over the years have been improved using partial matching\cite{wei2011improved}, bicliques\cite{bogdanov2011biclique}, sieve- in-the-middle\cite{canteaut2013sieve} etc. This approach requires fewer data and is an applied tool. The bicliques method can be used to reduce the total number
of computations, with the main focus of acceleration of exhaustive search. Bicliques\cite{jeong2012biclique} have been used for attacks on PRESENT, LED, KLEIN, HIGHT, Piccolo, TWINE, and LBlock. 

Merging the lists and dissection algorithm such as that for divide-and-conquer and rebound attacks find applications in ARMADILLO2\cite{abdelraheem2011cryptanalysis}, ECHO256\cite{naya2011improve}, JH42\cite{naya2011rebound},
Grøstl\cite{mendel2009rebound}, Klein,
AES-like, Sprout\cite{lallemand2015cryptanalysis}, and Ketje. Among the popular algorithm specific attack schemes, such as for PRESENT, the most effective approach as been multiple linear attacks using Sieving, forward and backward computations \cite{leander2011linear}.

\section{Memristor cryptography}

The majority of the cryptography works based on memristor circuits aim for low power and compact on-chip solution. Given that more and more devices are connected to internet, such solutions are ideally suitable for edge devices and can be considered within the class of lightweight cryptography solutions. 

\subsection{Chaotic systems}

The memristor due to this resistive switching behaviour forms as an excellent choice for building chaotic circuits\cite{muthuswamy2010implementing,zheng2018analysis}. The cryptographic application\cite{yang1997cryptography} of chaotic circuits range from that of random number generators to that for modelling dynamic systems. The state equations for chaotic systems can be parameterised using the memristor device, and offers an area efficient way to implement chaotic oscillators and circuits. The chaotic systems can be used to build chaotic encryptor and the chaotic decryptor for secure communication \cite{arafin2015survey}. The memristor based chaotic system also finds application in image encryption \cite{wang2018memristor}.

The use of random numbers are essential to ensure the difficulty of breaking a majority of cryptographic systems used today such as AES and RSA\cite{miller1982rsa}. The ability to guess the pseudo random numbers generated by conventional techniques within this algorithm can be a potential weakness that can be exploited by the attackers.  A chaotic random number generators\cite{corinto2016memristor} can overcome this issue by making it extremely difficult to predict the generated numbers.

\subsection{Physical unclonable functions}

Physical Unclonable Functions (PUFs) \cite{maes2016physically} from electronic circuits has a unique microstructure that results from the variability introduced during semiconductor manufacturing. The physical variability are unpredictable making it impossible to replicate its structure. PUFs are implemented using challenge–response authentication, where it evaluates the underlying microstructure. For a given stimulus (or challenge), the microstructure responds (or response) in unpredictable but in a precise manner. The challenge-response pair (CRP) does not reveal the device structure and hence is resistant to spoofing attacks.

The cryptographic keys can be also obtained using key extractor PUFs. The PUF hardware costs lower that a ROM based CRP that uses table of responses to the challenges. Even with same manufacturing process, the PUF from one device to another will be different, making it unclonable and difficult to compute unknown response. Without knowing all the physical properties its practically not possible to predict CRPs. This essentially means that PUFs are useful as unique signatures for edge devices, and is also useful for key generation and a source of randomness.

 The classical approaches to cryptography are often slow, energy consuming, and prone to various attacks. The physical unclonable functions \cite{suh2007physical} are hardware tokens that depends on the intrinsic behavior of memristor networks, and maps a challenge to a response. The public physical unclonable functions (PPUF) \cite{beckmann2009hardware,maes2010physically} is one of the prominent build using memristor crossbar arrays\cite{rajendran2012nano,gao2016emerging,mazady2015memristor,arafin2018memristors,uddin2018design}, and employs the non-idealities and characteristics of the memristor devices. Such PPUF can be used for multiple party security using keys, authentications, time stamping, and bit commitments.

Arguably, the most important aspect that makes memristor a suitable device for PUF is the ability to have randomness within a memristor network, making it a good building block for a complex physical system for extracting secret keys. In the past, the use of Ring Oscillator PUF (ROPUF)\cite{maiti2011improved}, Arbiter PUF (APUF)\cite{tajik2014physical}, and SRAM (Static Random Access Memory) PUF (SRAM PUF)\cite{garg2014design} used digital designs testing on FPGAs or ASICs. However, with scaling, these systems become unstable due to dependence of temperature and practical signal integrity issues. The use of nanoelectronic systems such as based on memristor could become popular for generating large challenge-response pairs, as they prove to be area efficient and provides an option for generating more stable PUF such as within crossbar networks\cite{zhang2018nanoscale,nili2018hardware}. The PUF design with memristors can be also extended to develop reconfigurable PUFs using different memristor network configurations, which help generalise the PUF approach to larger number of key exchange schemes\cite{gao2018efficient}.

\subsection{Hash functions}

The memristor crossbar can be used to build encrypted messaging systems, such as MemHash\cite{kvatinsky2018memristive,azriel2017towards}. In MemHash, a prefix and suffix is wrapped with original message. This message is further passed through a scrambler that is linear function of input bits, cycle count and a random value read from the crossbar array. This is used to generate an address and a value to write to the crossbar array. For the subsequent cycles, a differential read circuit is used to provide the input to the scrambler and for having a signature read-back.

The feasibility of using such hash functions in realistic systems requires further tests, as the quality of the crossbar devices can have an impact on how it is used in the hashing based algorithms. The interface circuits such as differential read block if inaccurate can have a significant impact on the performance of hash functions generated using the memhash systems. Nonetheless, this approach is useful as the technology matured and process related issues resolved.

\subsection{Open challenges}

\paragraph{Reliability issues}\added{There exists several open challenges in this area of work. The field of memristor cryptography is challenged by the reliability issues of memristor devices. The device and process variability in memristor crossbars is a useful aspect of the design of most memristor cryptographic systems. However, there are several practical reliability issues that are not usually accounted for in the design such as effect of aging, state variability, signal integrity and, electromagnetics issues.}

\paragraph{Variability}\added{The integration of the CMOS circuits with that of the memristor arrays in a cryptography chip is not a trivial task. Since the variability between a non-ideal memristor crossbar from one chip to another can be high, the process related variability that acts as an encoding signature expected from these devices would be hard to replicate under the effect of aging and temperature changes.
}

\paragraph{Architecture robustness}\added{The system integration and architecture for memristor cryptography is another open problem. While there are few class of architectures such as based on PUF and hash functions, they could be prone to side channel attacks when the designs are of small scale. Further, communication errors on-chip and off-chip  can be explored by adversaries to model the behaviour of the encryption scheme. }

\paragraph{Hardware acceleration}
\added{
The speed-up of traditional hardware implementations of cryptography algorithms is an on-going challenge for edge devices. There are dedicated cryptography chips that are incorporated as a co-processor in modern commercial edge devices. These co-processors uses digital gates and random number generators, which could be in future efficiently implemented with memristor threshold logic gates and chaotic generators. 
}

\paragraph{Neural cryptography}
\added{
Neural cryptography is an emerging field of study that is yet to be proven  to be a useful cryptography solution. In this approach, the human is kept out of the loop, while the encryption, decryption and adversaries are all neural network machines. Given that several different types of neural networks can be implemented with memristor crossbar arrays, it is possible to built and deploy the memristive neural cryptography solutions in the upcoming years.
}

\section{Discussions and concluding remarks}

The hardware security primitives are required to provide on-chip solutions that work at high speeds and provide additional layer of security as it is difficult to physically identify the on-chip circuits, and reduces the chance of the attacker to crack such systems. However, as a caution of note, the cryptanlysis for the memristor cryptographic systems is not a developed field. The understanding for dedicated attacks needs to be further investigated. The design ’risk’, low security margin, and fewer number of components in the memristor systems offers certain room for attacks. These systems are not yet fully tested for practical use.

The use of memristor circuits in traditional lightweight cyptographic methods for edge devices is an important and open problem. Since memristor networks can serve as associative memories they could be incorporated into different algorithmic cryptographic methods. The memristor circuits are also a good source for random key generation, that can make it useful for various traditional cryptographic methods.

The memristor behaviours are hard to replicate under realistic conditions. This makes it a good candidate for the PUFs. On the other hand, the impact of reliability, number of write-erase cycles, stability and the interconnect issues are not very well studied for practical use to build memristive cryptographic chips. The cryptanalysis over such hardware issues are nearly not studied at this stage in a practical context, and a substantial progress is required for memristive chips to be of realistic use in modern cryptography.

\section{Author contribution statement}
All contributions in the writing of this paper is done by A.P. James.
%\bibliographystyle{plain}
%\bibliography{ref}

\begin{thebibliography}{100}

\bibitem{abdelraheem2011cryptanalysis}
Mohamed~Ahmed Abdelraheem, C{\'e}line Blondeau, Mar{\'\i}a Naya-Plasencia,
  Marion Videau, and Erik Zenner.
\newblock Cryptanalysis of armadillo2.
\newblock In {\em International Conference on the Theory and Application of
  Cryptology and Information Security}, pages 308--326. Springer, 2011.

\bibitem{abraham2018case}
Isaac Abraham.
\newblock The case for rejecting the memristor as a fundamental circuit
  element.
\newblock {\em Scientific reports}, 8(1):10972, 2018.

\bibitem{adam20173}
Gina~C Adam, Brian~D Hoskins, Mirko Prezioso, Farnood Merrikh-Bayat, Bhaswar
  Chakrabarti, and Dmitri~B Strukov.
\newblock 3-d memristor crossbars for analog and neuromorphic computing
  applications.
\newblock {\em IEEE Transactions on Electron Devices}, 64(1):312--318, 2017.

\bibitem{albrecht2014block}
Martin~R Albrecht, Benedikt Driessen, Elif~Bilge Kavun, Gregor Leander,
  Christof Paar, and Tolga Yal{\c{c}}{\i}n.
\newblock Block ciphers--focus on the linear layer (feat. pride).
\newblock In {\em International Cryptology Conference}, pages 57--76. Springer,
  2014.

\bibitem{arafin2015survey}
Md~Tanvir Arafin, Carson Dunbar, Gang Qu, N~McDonald, and L~Yan.
\newblock A survey on memristor modeling and security applications.
\newblock In {\em Sixteenth International Symposium on Quality Electronic
  Design}, pages 440--447. IEEE, 2015.

\bibitem{arafin2018memristors}
Md~Tanvir Arafin and Gang Qu.
\newblock Memristors for secret sharing-based lightweight authentication.
\newblock {\em IEEE Transactions on Very Large Scale Integration (VLSI)
  Systems}, (99):1--13, 2018.

\bibitem{azriel2017towards}
Leonid Azriel and Shahar Kvatinsky.
\newblock Towards a memristive hardware secure hash function (memhash).
\newblock In {\em 2017 IEEE International Symposium on Hardware Oriented
  Security and Trust (HOST)}, pages 51--55. IEEE, 2017.

\bibitem{balasch2012compact}
Josep Balasch, Bari{\c{s}} Ege, Thomas Eisenbarth, Benoit G{\'e}rard, Zheng
  Gong, Tim G{\"u}neysu, Stefan Heyse, St{\'e}phanie Kerckhof, Fran{\c{c}}ois
  Koeune, Thomas Plos, et~al.
\newblock Compact implementation and performance evaluation of hash functions
  in attiny devices.
\newblock In {\em International Conference on Smart Card Research and Advanced
  Applications}, pages 158--172. Springer, 2012.

\bibitem{balasubramanian2018recent}
Kannan Balasubramanian.
\newblock Recent developments in cryptography: A survey.
\newblock In {\em Algorithmic Strategies for Solving Complex Problems in
  Cryptography}, pages 1--22. IGI Global, 2018.

\bibitem{beaulieu2015simon}
Ray Beaulieu, Stefan Treatman-Clark, Douglas Shors, Bryan Weeks, Jason Smith,
  and Louis Wingers.
\newblock The simon and speck lightweight block ciphers.
\newblock In {\em 2015 52nd ACM/EDAC/IEEE Design Automation Conference (DAC)},
  pages 1--6. IEEE, 2015.

\bibitem{beckmann2009hardware}
Nathan Beckmann and Miodrag Potkonjak.
\newblock Hardware-based public-key cryptography with public physically
  unclonable functions.
\newblock In {\em International Workshop on Information Hiding}, pages
  206--220. Springer, 2009.

\bibitem{biham1999miss}
Eli Biham, Alex Biryukov, and Adi Shamir.
\newblock Miss in the middle attacks on idea and khufu.
\newblock In {\em International Workshop on Fast Software Encryption}, pages
  124--138. Springer, 1999.

\bibitem{bogdanov2011biclique}
Andrey Bogdanov, Dmitry Khovratovich, and Christian Rechberger.
\newblock Biclique cryptanalysis of the full aes.
\newblock In {\em International Conference on the Theory and Application of
  Cryptology and Information Security}, pages 344--371. Springer, 2011.

\bibitem{borghoff2012prince}
Julia Borghoff, Anne Canteaut, Tim G{\"u}neysu, Elif~Bilge Kavun, Miroslav
  Knezevic, Lars~R Knudsen, Gregor Leander, Ventzislav Nikov, Christof Paar,
  Christian Rechberger, et~al.
\newblock Prince--a low-latency block cipher for pervasive computing
  applications.
\newblock In {\em International Conference on the Theory and Application of
  Cryptology and Information Security}, pages 208--225. Springer, 2012.

\bibitem{boura2014scrutinizing}
Christina Boura, Mar{\'\i}a Naya-Plasencia, and Valentin Suder.
\newblock Scrutinizing and improving impossible differential attacks:
  applications to clefia, camellia, lblock and simon.
\newblock In {\em International Conference on the Theory and Application of
  Cryptology and Information Security}, pages 179--199. Springer, 2014.

\bibitem{brier2002weierstrass}
Eric Brier and Marc Joye.
\newblock Weierstra elliptic curves and side-channel attacks.
\newblock In {\em International Workshop on Public Key Cryptography}, pages
  335--345. Springer, 2002.

\bibitem{buchmann2018postquantum}
Johannes Buchmann, Kristin Lauter, and Michele Mosca.
\newblock Postquantum cryptography, part 2.
\newblock {\em IEEE Security \& Privacy}, 16(5):12--13, 2018.

\bibitem{cache2007hacking}
Johnny Cache, Vincent Liu, and Joshua Wright.
\newblock {\em Hacking exposed wireless: wireless security secrets and
  solutions}.
\newblock McGraw-Hill, 2007.

\bibitem{canteaut2013sieve}
Anne Canteaut, Mar{\'\i}a Naya-Plasencia, and Bastien Vayssiere.
\newblock Sieve-in-the-middle: improved mitm attacks.
\newblock In {\em Advances in Cryptology--CRYPTO 2013}, pages 222--240.
  Springer, 2013.

\bibitem{cheng2008puffin}
Huiju Cheng, Howard~M Heys, and Cheng Wang.
\newblock Puffin: A novel compact block cipher targeted to embedded digital
  systems.
\newblock In {\em 2008 11th EUROMICRO Conference on Digital System Design
  Architectures, Methods and Tools}, pages 383--390. IEEE, 2008.

\bibitem{chua2018five}
L~Chua.
\newblock Five non-volatile memristor enigmas solved.
\newblock {\em Applied Physics A}, 124(8):563, 2018.

\bibitem{chua1971memristor}
Leon Chua.
\newblock Memristor-the missing circuit element.
\newblock {\em IEEE Transactions on circuit theory}, 18(5):507--519, 1971.

\bibitem{corinto2018memristor}
Fernando Corinto and Mauro Forti.
\newblock Memristor circuits: Pulse programming via invariant manifolds.
\newblock {\em IEEE Transactions on Circuits and Systems I: Regular Papers},
  65(4):1327--1339, 2018.

\bibitem{corinto2016memristor}
Fernando Corinto, V~Krulikovskyi, and Serhii~D Haliuk.
\newblock Memristor-based chaotic circuit for pseudo-random sequence
  generators.
\newblock In {\em 2016 18th Mediterranean Electrotechnical Conference
  (MELECON)}, pages 1--3. IEEE, 2016.

\bibitem{daemen2013design}
Joan Daemen and Vincent Rijmen.
\newblock {\em The design of Rijndael: AES-the advanced encryption standard}.
\newblock Springer Science \& Business Media, 2013.

\bibitem{damaj2018analysis}
Issam Damaj and Safaa Kasbah.
\newblock An analysis framework for hardware and software implementations with
  applications from cryptography.
\newblock {\em Computers and Electrical Engineering}, 69:572--584, 2018.

\bibitem{de2007high}
Guerric~Meurice de~Dormale and Jean-Jacques Quisquater.
\newblock High-speed hardware implementations of elliptic curve cryptography: A
  survey.
\newblock {\em Journal of systems architecture}, 53(2-3):72--84, 2007.

\bibitem{dorri2017blockchain}
Ali Dorri, Salil~S Kanhere, Raja Jurdak, and Praveen Gauravaram.
\newblock Blockchain for iot security and privacy: The case study of a smart
  home.
\newblock In {\em 2017 IEEE International Conference on Pervasive Computing and
  Communications Workshops (PerCom Workshops)}, pages 618--623. IEEE, 2017.

\bibitem{eisenbarth2007survey}
Thomas Eisenbarth, Sandeep Kumar, Christof Paar, Axel Poschmann, and Leif
  Uhsadel.
\newblock A survey of lightweight-cryptography implementations.
\newblock {\em IEEE Design \& Test of Computers}, 24(6):522--533, 2007.

\bibitem{el2015survey}
Nadia El~Mrabet, Jacques~JA Fournier, Louis Goubin, and Ronan Lashermes.
\newblock A survey of fault attacks in pairing based cryptography.
\newblock {\em Cryptography and Communications}, 7(1):185--205, 2015.

\bibitem{engels2010hummingbird}
Daniel Engels, Xinxin Fan, Guang Gong, Honggang Hu, and Eric~M Smith.
\newblock Hummingbird: ultra-lightweight cryptography for resource-constrained
  devices.
\newblock In {\em International Conference on Financial Cryptography and Data
  Security}, pages 3--18. Springer, 2010.

\bibitem{gao2018efficient}
Yansong Gao, Chenglu Jin, Jeeson Kim, Hussein Nili, Xiaolin Xu, Wayne Burleson,
  Omid Kavehei, Marten van Dijk, Damith~Chinthana Ranasinghe, and Ulrich
  R{\"u}hrmair.
\newblock Efficient erasable pufs from programmable logic and memristors.
\newblock {\em IACR Cryptology ePrint Archive}, 2018:358, 2018.

\bibitem{gao2016emerging}
Yansong Gao, Damith~C Ranasinghe, Said~F Al-Sarawi, Omid Kavehei, and Derek
  Abbott.
\newblock Emerging physical unclonable functions with nanotechnology.
\newblock {\em IEEE access}, 4:61--80, 2016.

\bibitem{garg2014design}
Achiranshu Garg and Tony~T Kim.
\newblock Design of sram puf with improved uniformity and reliability utilizing
  device aging effect.
\newblock In {\em 2014 IEEE International Symposium on Circuits and Systems
  (ISCAS)}, pages 1941--1944. IEEE, 2014.

\bibitem{gong2011klein}
Zheng Gong, Svetla Nikova, and Yee~Wei Law.
\newblock Klein: a new family of lightweight block ciphers.
\newblock In {\em International Workshop on Radio Frequency Identification:
  Security and Privacy Issues}, pages 1--18. Springer, 2011.

\bibitem{hankerson2011elliptic}
Darrel Hankerson and Alfred Menezes.
\newblock {\em Elliptic curve cryptography}.
\newblock Springer, 2011.

\bibitem{ho2011dynamical}
Yenpo Ho, Garng~M Huang, and Peng Li.
\newblock Dynamical properties and design analysis for nonvolatile memristor
  memories.
\newblock {\em IEEE Transactions on Circuits and Systems I: Regular Papers},
  58(4):724--736, 2011.

\bibitem{hong2006hight}
Deukjo Hong, Jaechul Sung, Seokhie Hong, Jongin Lim, Sangjin Lee, Bon-Seok Koo,
  Changhoon Lee, Donghoon Chang, Jesang Lee, Kitae Jeong, et~al.
\newblock Hight: A new block cipher suitable for low-resource device.
\newblock In {\em International Workshop on Cryptographic Hardware and Embedded
  Systems}, pages 46--59. Springer, 2006.

\bibitem{huff2008into}
Howard Huff.
\newblock {\em Into the nano era: Moore's law beyond planar silicon CMOS},
  volume 106.
\newblock Springer Science and Business Media, 2008.

\bibitem{irmanova2018neuron}
Aidana Irmanova and Alex~Pappachen James.
\newblock Neuron inspired data encoding memristive multi-level memory cell.
\newblock {\em Analog Integrated Circuits and Signal Processing},
  95(3):429--434, 2018.

\bibitem{izadi2009mibs}
Maryam Izadi, Babak Sadeghiyan, Seyed~Saeed Sadeghian, and Hossein~Arabnezhad
  Khanooki.
\newblock Mibs: a new lightweight block cipher.
\newblock In {\em International Conference on Cryptology and Network Security},
  pages 334--348. Springer, 2009.

\bibitem{jeong2012biclique}
Kitae Jeong, HyungChul Kang, Changhoon Lee, Jaechul Sung, and Seokhie Hong.
\newblock Biclique cryptanalysis of lightweight block ciphers present, piccolo
  and led.
\newblock {\em IACR Cryptology ePrint Archive}, 2012:621, 2012.

\bibitem{joglekar2009elusive}
Yogesh~N Joglekar and Stephen~J Wolf.
\newblock The elusive memristor: properties of basic electrical circuits.
\newblock {\em European Journal of Physics}, 30(4):661, 2009.

\bibitem{joy2011side}
G~Joy~Persial, M~Prabhu, and R~Shanmugalakshmi.
\newblock Side channel attack-survey.
\newblock {\em Int J Adva Sci Res Rev}, 1(4):54--57, 2011.

\bibitem{kahng2010scaling}
Andrew~B Kahng.
\newblock Scaling: More than moore's law.
\newblock {\em IEEE Design and Test of Computers}, 27(3):86--87, 2010.

\bibitem{karaklajic2013hardware}
Du{\v{s}}ko Karaklaji{\'c}, J{\"o}rn-Marc Schmidt, and Ingrid Verbauwhede.
\newblock Hardware designer's guide to fault attacks.
\newblock {\em IEEE Transactions on Very Large Scale Integration (VLSI)
  Systems}, 21(12):2295--2306, 2013.

\bibitem{karakocc2012impossible}
Ferhat Karako{\c{c}}, H{\"u}seyin Demirci, and A~Emre Harmanc{\i}.
\newblock Impossible differential cryptanalysis of reduced-round lblock.
\newblock In {\em IFIP International Workshop on Information Security Theory
  and Practice}, pages 179--188. Springer, 2012.

\bibitem{katz1996handbook}
Jonathan Katz, Alfred~J Menezes, Paul~C Van~Oorschot, and Scott~A Vanstone.
\newblock {\em Handbook of applied cryptography}.
\newblock CRC press, 1996.

\bibitem{kim2003impossible}
Jongsung Kim, Seokhie Hong, Jaechul Sung, Sangjin Lee, Jongin Lim, and Soohak
  Sung.
\newblock Impossible differential cryptanalysis for block cipher structures.
\newblock In {\em International Conference on Cryptology in India}, pages
  82--96. Springer, 2003.

\bibitem{kim2011functional}
Kuk-Hwan Kim, Siddharth Gaba, Dana Wheeler, Jose~M Cruz-Albrecht, Tahir
  Hussain, Narayan Srinivasa, and Wei Lu.
\newblock A functional hybrid memristor crossbar-array/cmos system for data
  storage and neuromorphic applications.
\newblock {\em Nano letters}, 12(1):389--395, 2011.

\bibitem{knudsen2010printcipher}
Lars Knudsen, Gregor Leander, Axel Poschmann, and Matthew~JB Robshaw.
\newblock Printcipher: a block cipher for ic-printing.
\newblock In {\em International Workshop on Cryptographic Hardware and Embedded
  Systems}, pages 16--32. Springer, 2010.

\bibitem{knudsen1994truncated}
Lars~R Knudsen.
\newblock Truncated and higher order differentials.
\newblock In {\em International Workshop on Fast Software Encryption}, pages
  196--211. Springer, 1994.

\bibitem{kolay2014khudra}
Souvik Kolay and Debdeep Mukhopadhyay.
\newblock Khudra: a new lightweight block cipher for fpgas.
\newblock In {\em International Conference on Security, Privacy, and Applied
  Cryptography Engineering}, pages 126--145. Springer, 2014.

\bibitem{krestinskaya2019memristive}
Olga Krestinskaya, Aidana Irmanova, and Alex~Pappachen James.
\newblock Memristive non-idealities: Is there any practical implications for
  designing neural network chips?
\newblock In {\em IEEE International Symposium on Circuits and Systems}, 2019.

\bibitem{krestinskaya2019neuro}
Olga Krestinskaya, Alex~Pappachen James, and Leon~O Chua.
\newblock Neuro-memristive circuits for edge computing: A review.
\newblock {\em IEEE Transactions on Neural Networks and Learning Systems},
  2019.

\bibitem{krestinskaya2018analog}
Olga Krestinskaya, Khaled~Nabil Salama, and Alex~Pappachen James.
\newblock Analog backpropagation learning circuits for memristive crossbar
  neural networks.
\newblock In {\em 2018 IEEE International Symposium on Circuits and Systems
  (ISCAS)}, pages 1--5. IEEE, 2018.

\bibitem{kuekes2005crossbar}
Philip~J Kuekes, Duncan~R Stewart, and R~Stanley Williams.
\newblock The crossbar latch: Logic value storage, restoration, and inversion
  in crossbar circuits.
\newblock {\em Journal of Applied Physics}, 97(3):034301, 2005.

\bibitem{kvatinsky2018memristive}
Shahar Kvatinsky and Leonid Azriel.
\newblock Memristive security hash function, November~1 2018.
\newblock US Patent App. 15/965,924.

\bibitem{lallemand2015cryptanalysis}
Virginie Lallemand and Mar{\'\i}a Naya-Plasencia.
\newblock Cryptanalysis of full sprout.
\newblock In {\em Annual Cryptology Conference}, pages 663--682. Springer,
  2015.

\bibitem{leander2011linear}
Gregor Leander.
\newblock On linear hulls, statistical saturation attacks, present and a
  cryptanalysis of puffin.
\newblock In {\em Annual International Conference on the Theory and
  Applications of Cryptographic Techniques}, pages 303--322. Springer, 2011.

\bibitem{li2018analogue}
Can Li, Miao Hu, Yunning Li, Hao Jiang, Ning Ge, Eric Montgomery, Jiaming
  Zhang, Wenhao Song, Noraica D{\'a}vila, Catherine~E Graves, et~al.
\newblock Analogue signal and image processing with large memristor crossbars.
\newblock {\em Nature Electronics}, 1(1):52, 2018.

\bibitem{lim2005mcrypton}
Chae~Hoon Lim and Tymur Korkishko.
\newblock mcrypton--a lightweight block cipher for security of low-cost rfid
  tags and sensors.
\newblock In {\em International Workshop on Information Security Applications},
  pages 243--258. Springer, 2005.

\bibitem{lu2011two}
Wei Lu, Kuk-Hwan Kim, Ting Chang, and Siddharth Gaba.
\newblock Two-terminal resistive switches (memristors) for memory and logic
  applications.
\newblock In {\em Proceedings of the 16th Asia and South Pacific Design
  Automation Conference}, pages 217--223. IEEE Press, 2011.

\bibitem{maes2016physically}
Roel Maes.
\newblock {\em Physically unclonable functions}.
\newblock Springer, 2016.

\bibitem{maes2010physically}
Roel Maes and Ingrid Verbauwhede.
\newblock Physically unclonable functions: A study on the state of the art and
  future research directions.
\newblock In {\em Towards Hardware-Intrinsic Security}, pages 3--37. Springer,
  2010.

\bibitem{maiti2011improved}
Abhranil Maiti and Patrick Schaumont.
\newblock Improved ring oscillator puf: An fpga-friendly secure primitive.
\newblock {\em Journal of cryptology}, 24(2):375--397, 2011.

\bibitem{mala2010improved}
Hamid Mala, Mohammad Dakhilalian, Vincent Rijmen, and Mahmoud Modarres-Hashemi.
\newblock Improved impossible differential cryptanalysis of 7-round aes-128.
\newblock In {\em International Conference on Cryptology in India}, pages
  282--291. Springer, 2010.

\bibitem{mala2011impossible}
Hamid Mala, Mohammad Dakhilalian, and Mohsen Shakiba.
\newblock Impossible differential attacks on 13-round clefia-128.
\newblock {\em Journal of Computer Science and Technology}, 26(4):744--750,
  2011.

\bibitem{mazady2015memristor}
Anas Mazady, Md~Tauhidur Rahman, Domenic Forte, and Mehdi Anwar.
\newblock Memristor puf—a security primitive: Theory and experiment.
\newblock {\em IEEE Journal on Emerging and Selected Topics in Circuits and
  Systems}, 5(2):222--229, 2015.

\bibitem{mendel2009rebound}
Florian Mendel, Christian Rechberger, Martin Schl{\"a}ffer, and S{\o}ren~S
  Thomsen.
\newblock The rebound attack: Cryptanalysis of reduced whirlpool and gr{\o}stl.
\newblock In {\em International Workshop on Fast Software Encryption}, pages
  260--276. Springer, 2009.

\bibitem{miller1982rsa}
William~J Miller and Nick~G Trbovich.
\newblock Rsa public-key data encryption system having large random prime
  number generating microprocessor or the like, September~28 1982.
\newblock US Patent 4,351,982.

\bibitem{moosavi2015sea}
Sanaz~Rahimi Moosavi, Tuan~Nguyen Gia, Amir-Mohammad Rahmani, Ethiopia
  Nigussie, Seppo Virtanen, Jouni Isoaho, and Hannu Tenhunen.
\newblock Sea: a secure and efficient authentication and authorization
  architecture for iot-based healthcare using smart gateways.
\newblock {\em Procedia Computer Science}, 52:452--459, 2015.

\bibitem{mouttet2008proposal}
Blaise Mouttet.
\newblock Proposal for memristor crossbar design and applications.
\newblock In {\em Memristors and Memristive Systems Symposium, UC Berkeley},
  2008.

\bibitem{mouttet2007programmable}
Blaise~Laurent Mouttet.
\newblock Programmable crossbar signal processor, November~27 2007.
\newblock US Patent 7,302,513.

\bibitem{muthuswamy2010implementing}
Bharathwaj Muthuswamy.
\newblock Implementing memristor based chaotic circuits.
\newblock {\em International Journal of Bifurcation and Chaos},
  20(05):1335--1350, 2010.

\bibitem{naya2011improve}
Mar{\'\i}a Naya-Plasencia.
\newblock How to improve rebound attacks.
\newblock In {\em Annual Cryptology Conference}, pages 188--205. Springer,
  2011.

\bibitem{naya2011rebound}
Mar{\'\i}a Naya-Plasencia, Deniz Toz, and Kerem Varici.
\newblock Rebound attack on jh42.
\newblock In {\em International Conference on the Theory and Application of
  Cryptology and Information Security}, pages 252--269. Springer, 2011.

\bibitem{nili2018hardware}
Hussein Nili, Gina~C Adam, Brian Hoskins, Mirko Prezioso, Jeeson Kim, M~Reza
  Mahmoodi, Farnood~Merrikh Bayat, Omid Kavehei, and Dmitri~B Strukov.
\newblock Hardware-intrinsic security primitives enabled by analogue state and
  nonlinear conductance variations in integrated memristors.
\newblock {\em Nature Electronics}, 1(3):197, 2018.

\bibitem{nishimura1990probability}
Kazuo Nishimura and Masaaki Sibuya.
\newblock Probability to meet in the middle.
\newblock {\em Journal of Cryptology}, 2(1):13--22, 1990.

\bibitem{ometov2016feasibility}
Aleksandr Ometov, Pavel Masek, Lukas Malina, Roman Florea, Jiri Hosek, Sergey
  Andreev, Jan Hajny, Jussi Niutanen, and Yevgeni Koucheryavy.
\newblock Feasibility characterization of cryptographic primitives for
  constrained (wearable) iot devices.
\newblock In {\em 2016 IEEE International Conference on Pervasive Computing and
  Communication Workshops (PerCom Workshops)}, pages 1--6. IEEE, 2016.

\bibitem{panasenko2011lightweight}
Sergey Panasenko and Sergey Smagin.
\newblock Lightweight cryptography: Underlying principles and approaches.
\newblock {\em International Journal of Computer Theory and Engineering},
  3(4):516, 2011.

\bibitem{pantelopoulos2010survey}
Alexandros Pantelopoulos and Nikolaos~G Bourbakis.
\newblock A survey on wearable sensor-based systems for health monitoring and
  prognosis.
\newblock {\em IEEE Transactions on Systems, Man, and Cybernetics, Part C
  (Applications and Reviews)}, 40(1):1--12, 2010.

\bibitem{poschmann2009lightweight}
Axel~York Poschmann.
\newblock Lightweight cryptography: cryptographic engineering for a pervasive
  world.
\newblock In {\em PH. D. THESIS}. Citeseer, 2009.

\bibitem{rajagopalan2012survey}
Sundararaman Rajagopalan, Rengarajan Amirtharajan, Har~Narayan Upadhyay, and
  John Bosco~Balaguru Rayappan.
\newblock Survey and analysis of hardware cryptographic and steganographic
  systems on fpga.
\newblock {\em Journal of Applied Sciences}, 12(3):201, 2012.

\bibitem{rajendran2012nano}
Jeyavijayan Rajendran, Garrett~S Rose, Ramesh Karri, and Miodrag Potkonjak.
\newblock Nano-ppuf: A memristor-based security primitive.
\newblock In {\em 2012 IEEE Computer Society Annual Symposium on VLSI}, pages
  84--87. IEEE, 2012.

\bibitem{schneier2000self}
Bruce Schneier.
\newblock A self-study course in block-cipher cryptanalysis.
\newblock {\em Cryptologia}, 24(1):18--33, 2000.

\bibitem{sergienko2018quantum}
Alexander~V Sergienko.
\newblock {\em Quantum communications and cryptography}.
\newblock CRC press, 2018.

\bibitem{shibutani2011piccolo}
Kyoji Shibutani, Takanori Isobe, Harunaga Hiwatari, Atsushi Mitsuda, Toru
  Akishita, and Taizo Shirai.
\newblock Piccolo: an ultra-lightweight blockcipher.
\newblock In {\em International Workshop on Cryptographic Hardware and Embedded
  Systems}, pages 342--357. Springer, 2011.

\bibitem{shirai2007128}
Taizo Shirai, Kyoji Shibutani, Toru Akishita, Shiho Moriai, and Tetsu Iwata.
\newblock The 128-bit blockcipher clefia.
\newblock In {\em International workshop on fast software encryption}, pages
  181--195. Springer, 2007.

\bibitem{stamp2007applied}
Mark Stamp and Richard~M Low.
\newblock {\em Applied cryptanalysis: breaking ciphers in the real world}.
\newblock John Wiley \& Sons, 2007.

\bibitem{stathopoulos2017multibit}
Spyros Stathopoulos, Ali Khiat, Maria Trapatseli, Simone Cortese, Alexantrou
  Serb, Ilia Valov, and Themis Prodromakis.
\newblock Multibit memory operation of metal-oxide bi-layer memristors.
\newblock {\em Scientific reports}, 7(1):17532, 2017.

\bibitem{struik2013aead}
Ren{\'e} Struik.
\newblock Aead ciphers for highly constrained networks.
\newblock DIAC, 2013.

\bibitem{strukov2008missing}
Dmitri~B Strukov, Gregory~S Snider, Duncan~R Stewart, and R~Stanley Williams.
\newblock The missing memristor found.
\newblock {\em nature}, 453(7191):80, 2008.

\bibitem{suh2007physical}
G~Edward Suh and Srinivas Devadas.
\newblock Physical unclonable functions for device authentication and secret
  key generation.
\newblock In {\em 2007 44th ACM/IEEE Design Automation Conference}, pages
  9--14. IEEE, 2007.

\bibitem{sun2015links}
Bing Sun, Zhiqiang Liu, Vincent Rijmen, Ruilin Li, Lei Cheng, Qingju Wang, Hoda
  Alkhzaimi, and Chao Li.
\newblock Links among impossible differential, integral and zero correlation
  linear cryptanalysis.
\newblock In {\em Annual Cryptology Conference}, pages 95--115. Springer, 2015.

\bibitem{tajik2014physical}
Shahin Tajik, Enrico Dietz, Sven Frohmann, Jean-Pierre Seifert, Dmitry
  Nedospasov, Clemens Helfmeier, Christian Boit, and Helmar Dittrich.
\newblock Physical characterization of arbiter pufs.
\newblock In {\em International Workshop on Cryptographic Hardware and Embedded
  Systems}, pages 493--509. Springer, 2014.

\bibitem{tomoyasu2012twine}
Suzaki Tomoyasu.
\newblock Twine: A lightweight block cipher for multiple platforms.
\newblock In {\em Selected Areas in Cryptography}, volume 7707. Springer Berlin
  Heidelberg, 2012.

\bibitem{tripathy2013esf}
Somanath Tripathy.
\newblock Esf: an efficient security framework for wireless sensor network.
\newblock {\em International Journal of Communication Networks and Distributed
  Systems}, 10(2):176--194, 2013.

\bibitem{uddin2018design}
Mesbah Uddin, MD~Majumder, Karsten Beckmann, Harika Manem, Zahiruddin Alamgir,
  Nathaniel~C Cady, and Garrett~S Rose.
\newblock Design considerations for memristive crossbar physical unclonable
  functions.
\newblock {\em ACM Journal on Emerging Technologies in Computing Systems
  (JETC)}, 14(1):2, 2018.

\bibitem{vongehr2015missing}
Sascha Vongehr and Xiangkang Meng.
\newblock The missing memristor has not been found.
\newblock {\em Scientific reports}, 5:11657, 2015.

\bibitem{vontobel2009writing}
Pascal~O Vontobel, Warren Robinett, Philip~J Kuekes, Duncan~R Stewart, Joseph
  Straznicky, and R~Stanley Williams.
\newblock Writing to and reading from a nano-scale crossbar memory based on
  memristors.
\newblock {\em Nanotechnology}, 20(42):425204, 2009.

\bibitem{wang2018memristor}
B~Wang, FC~Zou, and J~Cheng.
\newblock A memristor-based chaotic system and its application in image
  encryption.
\newblock {\em Optik}, 154:538--544, 2018.

\bibitem{wang2014cryptanalysis}
Qingju Wang, Zhiqiang Liu, Kerem Var{\i}c{\i}, Yu~Sasaki, Vincent Rijmen, and
  Yosuke Todo.
\newblock Cryptanalysis of reduced-round simon32 and simon48.
\newblock In {\em International Conference in Cryptology in India}, pages
  143--160. Springer, 2014.

\bibitem{warren2006social}
Matthew Warren and Shona Leitch.
\newblock Social engineering and its impact via the internet.
\newblock In {\em Proceedings of the 4th Australian Information Security
  Management Conference}, pages 184--189. Australian Information Security
  Management, 2006.

\bibitem{wei2011improved}
Lei Wei, Christian Rechberger, Jian Guo, Hongjun Wu, Huaxiong Wang, and San
  Ling.
\newblock Improved meet-in-the-middle cryptanalysis of ktantan (poster).
\newblock In {\em Australasian Conference on Information Security and Privacy},
  pages 433--438. Springer, 2011.

\bibitem{williams2017s1}
R~Stanley Williams.
\newblock What's next? the end of moore's law.
\newblock {\em Computing in Science and Engineering}, 19(2):7--13, 2017.

\bibitem{wu2007impossible}
Wen-Ling Wu, Wen-Tao Zhang, and Deng-Guo Feng.
\newblock Impossible differential cryptanalysis of reduced-round aria and
  camellia.
\newblock {\em Journal of Computer Science and Technology}, 22(3):449--456,
  2007.

\bibitem{wu2011lblock}
Wenling Wu and Lei Zhang.
\newblock Lblock: a lightweight block cipher.
\newblock In {\em International Conference on Applied Cryptography and Network
  Security}, pages 327--344. Springer, 2011.

\bibitem{yang1997cryptography}
Tao Yang, Chai~Wah Wu, and Leon~O Chua.
\newblock Cryptography based on chaotic systems.
\newblock {\em IEEE Transactions on Circuits and Systems I: Fundamental Theory
  and Applications}, 44(5):469--472, 1997.

\bibitem{zhang2018nanoscale}
R~Zhang, H~Jiang, ZR~Wang, P~Lin, Y~Zhuo, D~Holcomb, DH~Zhang, JJ~Yang, and
  Q~Xia.
\newblock Nanoscale diffusive memristor crossbars as physical unclonable
  functions.
\newblock {\em Nanoscale}, 10(6):2721--2726, 2018.

\bibitem{zhang2018neuromorphic}
Xinjiang Zhang, Anping Huang, Qi~Hu, Zhisong Xiao, and Paul~K Chu.
\newblock Neuromorphic computing with memristor crossbar.
\newblock {\em physica status solidi (a)}, 215(13):1700875, 2018.

\bibitem{zhang2014iot}
Zhi-Kai Zhang, Michael Cheng~Yi Cho, Chia-Wei Wang, Chia-Wei Hsu, Chong-Kuan
  Chen, and Shiuhpyng Shieh.
\newblock Iot security: ongoing challenges and research opportunities.
\newblock In {\em 2014 IEEE 7th international conference on service-oriented
  computing and applications}, pages 230--234. IEEE, 2014.

\bibitem{zheng2018analysis}
Ciyan Zheng, Herbert~HC Iu, Tyrone Fernando, Dongsheng Yu, Hengdao Guo, and
  Jason~K Eshraghian.
\newblock Analysis and generation of chaos using compositely connected coupled
  memristors.
\newblock {\em Chaos: An Interdisciplinary Journal of Nonlinear Science},
  28(6):063115, 2018.

\bibitem{zhou2005side}
YongBin Zhou and DengGuo Feng.
\newblock Side-channel attacks: Ten years after its publication and the impacts
  on cryptographic module security testing.
\newblock {\em IACR Cryptology ePrint Archive}, 2005:388, 2005.

\end{thebibliography}

\end{document}